\documentclass[%
 reprint,
 amsmath,amssymb,
pra,
]{revtex4-1}
\usepackage{color}
\usepackage{graphicx}% Include figure files
\usepackage{dcolumn}% Align table columns on decimal point
\usepackage{bm}% bold math
\begin{document}

\preprint{APS/123-QED}

\title{Ground state cooling of micromechanical oscillators in 
	the unresolved sideband regime induced by a quantum well}

\author{Bijita Sarma}
 %\altaffiliation[Also at ]{Department of Physics, Indian Institute of Technology  Guwahati, Guwahati-781039, Assam, India}
 \email{s.bijita@iitg.ernet.in}
\author{Amarendra K. Sarma}%
 \email{aksarma@iitg.ernet.in}
\affiliation{%
 Department of Physics, Indian Institute of Technology  Guwahati, Guwahati-781039, Assam, India}

%\collaboration{MUSO Collaboration}

%\affiliation{
%\affiliation{
%\affiliation{%

\date{\today}% It is always \today, today,

\begin{abstract}
We theoretically demonstrate the ground state cooling of a  mechanical 
oscillator in an optomechanical cavity in presence of a 
quantum well, in the unresolved-sideband regime. Due to the presence of the quantum well, the cavity  response 
gets modified and leads to asymmetric heating and cooling 
processes. The cooling rate of the mechanical resonator can potentially  be 
enhanced  by tuning the cavity-field 
detuning. It is demonstrated that, even when the cavity is in the unresolved-sideband 
regime, the effective interaction of the exciton and mechanical 
modes can bring the system back to effective resolved-sideband regime. Time 
evolution of the mean phonon number in the mechanical resonator is 
studied using the quantum master equation. The average phonon occupancy in the 
mechanical resonator tends to zero with time, 
exhibiting dynamic controllability of cavity dissipation.
\begin{description}
\item[PACS numbers]
\verb|42.50.Wk, 07.10.Cm, 42.50.Lc|
\end{description}
\end{abstract}

\pacs{Valid PACS appear here}% PACS, the Physics and Astronomy
\maketitle

%\tableofcontents

\section{\label{sec:level1}INTRODUCTION}
In recent years, the field of cavity 
optomechanics has been subjected to rapid exploration in terms of both 	theory 
and experiment [1-6]. The basic principle of coupling through radiation pressure 
force, as predicted
in [7-9] can lead to Kerr type 
nonlinearity between the optical and mechanical modes [10-11]. The 
optomechanical interaction between optical and mechanical modes is not only a 
tool for readout of mechanical motion [12], but it also triggers 
the possibility of observing fundamental quantum effects in mesoscopic 	systems. 
Owing to the inherent nonlinearity, cavity optomechanics is a strong candidate 
for futuristic aspects of achievement of standard 
quantum limit [13], continuous variable entanglement of optical and mechanical 
modes [14], nonclassical state generation [15], quantum state transfer between 
different modes [16], optomechanically induced 
transparency [17], quantum nondemolition measurements [18] etc. 

For observing quantum effects in optomechanical systems, ground state cooling of 
the mechanical oscillator is an essential condition [19, 20]. In an optomechanical 
system, the light scattered from 
the movable end mirror gives rise to Stokes and anti-Stokes sidebands. During 
the Stokes process, the mirror absorbs a quantum of energy from the cavity 
optical field, leading to heating of the mirror; whereas during the anti-Stokes 
process, the cavity field absorbs energy from the 
mirror resulting in cooling of the mirror. To obtain an effective cooling of the 
mechanical mirror, cooling rate should be higher than the heating rate. 
Therefore, in analogy to the laser cooling of ions in the strong 
binding regime [21], conventional cavity cooling of mechanical oscillators 
requires the condition of resolved-sideband regime, where the cavity mode decay 
rate is lower than the mechanical oscillator 
resonance frequency [19, 22].  However, in practical situations, for typical mechanical oscillators of frequency 
in the range of kHz-MHz; fulfilling this 
condition is a challenging task, that poses serious 
constraints experimentally. To relax this requirement, few different approaches 
have been suggested such as cooling using dissipative coupling [23-24], coupling 
with high-Q auxiliary cavity [25], hybrid atom-
optomechanical systems [26], optomechanically induced transparency [27]. Here, in this paper, we consider the cavity 
cooling of a micromechanical mirror, with a quantum well (QW) having lower 
exciton decay rate, placed inside the optomechanical cavity. 
This type of solid-state systems has their own advantages over atomic cavity 
systems. Engineerable emission frequency, fixed position and potential for 
integration with cavities and waveguides using developed 
semiconductor fabrication techniques [28] make them unique tool for exploring 
optomechanics further [6, 29]. The same type of systems has been studied in the 
context of nonlinear effects like bistability and 
squeezing of the output field [30]. 
It is also predicted in such a system that 
the interaction between the exciton and mechanical modes through the cavity 
field may lead to entanglement between the two [31]. We explore 
the aspect of ground state cooling of the mechanical oscillator in the 	
unresolved-sideband regime. 
%%%%%%%%%%%%%% FIGURE 1 %%%%%%%%%%%
\begin {figure}[t]
\begin {center}
\includegraphics [width =8 cm]{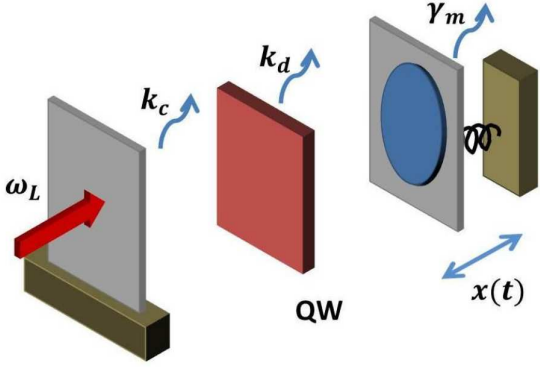}
\caption {(Color online) An optomechanical cavity containing a Quantum Well (QW) placed at the antinode of the cavity field.}
\label {fig1}
\end{center}
\end{figure}
{In this regime, by coupling the mechanical 
oscillator to a quantum well placed at the anti-node of the cavity field, 
one can modify the noise spectrum. This illustrates mode structuring around the 
sidebands, due to the inhibition of the heating process while enhancing the 
cooling process through quantum interference. In order to study the cooling process in the mechanical oscillator, we use an effective exciton-phonon interaction model that is analogous to dark mode formation, which has been studied extensively in optomechanics, in connection to state transfer protocols[32].  %In 
	%passing, it is worthwhile to note that recently various alternative methods for 
	%overcoming thermal motion, other than cooling a mechanical 
	%oscillator to its motional ground state, has been proposed [32]. One such 
	%approach exploits the use of a mechanically dark mode. 
	The dark mode in optomechanics is similar to 
	coherent-population trapped state or the dark state in atomic physics 
	[33]. The similarity to the so-called stimulated Raman adiabatic passage 
	(STIRAP) is also notable. In fact, STIRAP protocols are also studied extensively in 
	the context of state transfer in optomechanical and 
	electromechanical systems [32].} The paper is 
organized as follows. In Section II we describe the Hamiltonian of the 	system 
and derive the quantum Langevin equations for the system operators. Section III 
is devoted to the analysis of cooling of the 
mechanical oscillator, followed by conclusion of our work in Section IV.
\section{\label{sec:level1}MODEL AND THEORY}
We consider an optomechanical cavity containing a QW as shown schematically in Fig. 1.	The movable end mirror has resonance frequency $\omega_m$, effective mass $m$ and decay rate $\gamma_m$. The cavity is driven by an intense pump laser of frequency $\omega_l$, which exerts a radiation 
pressure force on the movable end mirror. The Hamiltonian of the 
system is given by (in the unit of $\hbar =1$)
\begin{align} \label{eq1}
H=H_{free}+H_{o-m}+H_{o-d}+H_{drive} 
\end{align} 
The first term $H_{free}$ in Eq. (1) describes the free Hamiltonian of the system, given by $H_{free}=\omega_cc^\dagger c+\omega_dd^\dagger d+{\omega}_mb^\dagger b$, 
where, $\omega_c$, $\omega_d$ and $\omega_m$ 
are the resonance frequencies of the cavity optical field, the QW excitons and the mechanical oscillator respectively. Since we are dealing with quasi-resonant coherent excitation, higher lying exciton states of the QW are neglected. The optomechanical interaction between the cavity mode and the mechanical oscillator is described by the second term, 
%%%%%%%%%%%%%% FIGURE 2 %%%%%%%%%%%
\begin {figure}[t]
\begin {center}
\includegraphics [width =8.5 cm]{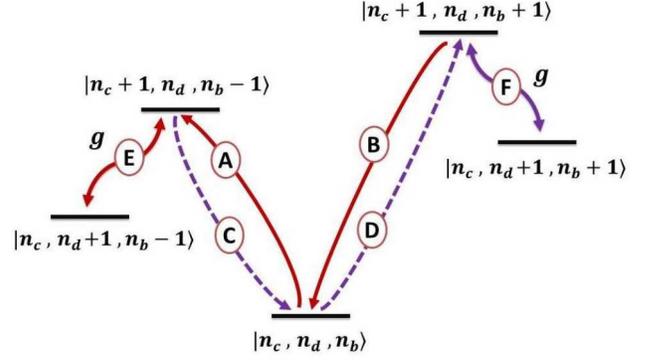}
\caption {(Color online) Level diagram for the linearized Hamiltonian [Eq. (4)] and different coupling routes in displaced frame where $|n_c,n_d,n_b\rangle$ denotes the state with $n_c$ number of photons, $n_d$ number of excitons and $n_b$ number of phonons. The solid, red curve $A (B)$ denote the cooling process due to energy swapping (counter-rotating-wave interaction). The dashed, purple curve $C (D)$ denote the heating process due to energy swapping (quantum backaction). The energy swapping due to exciton-cavity coupling is denoted by solid curves $E$ and $F$.}
\label {fig2}
\end{center}
\end{figure}
$H_{o-m}=g_{OM}c^\dagger c\left(b^\dagger+b\right)$. Here $g_{OM}$ is the 
single-photon optomechanical coupling strength between the cavity field 
and the mechanical oscillator. The third term in Eq. (1) accounts for the coupling between the cavity mode and the exciton mode in the QW, given by 
$H_{o-d}=g\left(c^\dagger d+d^\dagger c\right)$, with interaction strength  
$g$. The last term represents the pump laser driving, given by 
$H_{drive}={\varepsilon}_p\left(c^\dagger e^{-i\omega_lt}+ce^{i\omega_lt}\right)$, with pump laser frequency $\omega _l$ and amplitude 
${\varepsilon}_p=\sqrt{\frac{k_cP_{in}}{\hbar \omega_l}}$; $P_{in}$ and 
$k_c$ being the input power and cavity decay rate respectively. In the frame 
rotating with the input laser frequency $\omega_l$, one can obtain the 
Hamiltonian of the system as follows: 
\begin{align}\label{eq2} 
\nonumber
H =-\Delta_cc^\dagger c-\Delta_dd^\dagger d+\omega_mb^\dagger b+g_{OM}c^\dagger c\left(b^\dagger+b\right)\\ 
+g\left(c^\dagger d+d^\dagger c\right)+{\varepsilon}_p\left(c^\dagger +c\right)
\end{align}      
where, ${\mathrm{\Delta }}_c={\omega }_l-{\omega }_c$ and ${\mathrm{\Delta 
	}}_d={\omega}_l-{\omega }_d$ are the detunings of the cavity mode and the 
	exciton mode respectively. The time evolution of the system operators are given by nonlinear 
	Heisenberg-Langevin equations:
\begin{subequations}
\begin{align}\label{eq3}
\nonumber
\dot{c}=\left(i{\Delta}_c-\frac{k_c}{2}\right)c-ig_{OM}c\left(b^{\dagger}+b\right)-\\
%\nonumber
igd-i{\varepsilon }_p-\sqrt{k_c}c_{in}(t)\\
%\nonumber
\dot{d}=\left(i{\mathrm{\Delta}}_d-\frac{k_d}{2}\right)d-igc-\sqrt{k_d}d_{in}(t)\\
%\nonumber
\dot{b}=\left(-i{\omega}_m-\frac{{\gamma }_m}{2}\right)b-ig_{OM}c^{\dagger}c-\sqrt{{\gamma }_m}b_{in}(t)
\end{align}
\end{subequations}
%%%%%%%%%%%%%% FIGURE 3 %%%%%%%%%%%
\begin {figure}[t]
\begin {center}
\includegraphics [width =9 cm]{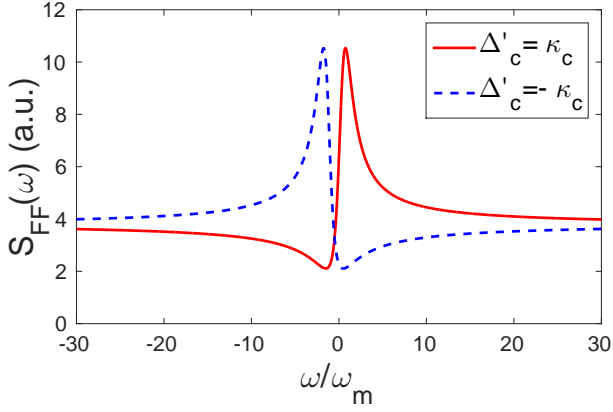}
\end{center}
\caption {(Color online) Plot of noise spectrum $S_{FF}$ vs. normalized frequency $\omega/\omega_m$ for $\Delta_c'=k_c$  (red solid),  $\Delta_c'=-k_c$ (blue dashed). Other parameters are $\Delta_d=0.5\omega_m$, $\gamma_m=10^{-5} \omega_m$, $k_c=10^4 \omega_m$, $k_d=\omega_m$, $g=100\omega_m$, $G=50\omega_m$  and $n_{th}=10^4$.}
\label {fig3}
\end{figure}
where, $k_c$, $k_d$ and ${\gamma }_m$ are the decay rates of the optical mode, 
exciton mode and the mechanical mode respectively. $c_{in}$, $d_{in}$ and 
$b_{in}$ are the corresponding input vacuum noise operators with zero mean value and nonzero correlation functions given by: 
$\left\langle c_{in}\left(t\right){c_{in}}^{\dagger}\left(t'\right)\right\rangle=\delta 
(t-t'), 
\left\langle 
d_{in}\left(t\right){d_{in}}^{\dagger}\left(t'\right)\right\rangle=\delta 
(t-t'), 
\left\langle b_{in}\left(t\right){b_{in}}^{\dagger 
}\left(t'\right)\right\rangle=(n_{th}+1)\delta (t-t'), 
\left\langle b_{in}^{\dagger 
	}\left(t\right)b_{in}\left(t'\right)\right\rangle =n_{th}\delta (t-t')
$ [35]. $n_{th}$ is the environmental thermal phonon number given by, 
		$n_{th}={{\mathrm{[exp} \left(\frac{\hslash {\omega }_m}{k_BT}\right)\ 
			}-1]}^{-1}$, where $k_B$ is Boltzmann constant and $T$ is the 
		environmental temperature. 
\section{\label{sec:level1}COOLING OF THE MECHANICAL OSCILLATOR} 
Assuming a strong driving condition, Eqs. (3a)-(3c) can be linearized 
around the steady state mean values by expressing the system operators to be 
comprising of the mean value, $O_S$, and a small fluctuating term, $\delta O$: 
$O=O_S+\delta O$. Then the linearized Hamiltonian of the system is given by:
\begin{align}\label{eq4}
\nonumber
H_L&=-{\mathrm{\Delta }'}_c {\delta c}^{\dagger}\delta c-{\mathrm{\Delta 
	}}_d{\delta d}^{\dagger }\delta d+{\omega }_m{\delta b}^{\dagger 
}\delta 
b +
G\left({\delta c}^{\dagger }+ \delta 
c\right)\\
&\left({\delta b}^{\dagger }+\delta b\right)+\ 
g\left({\delta 
	c}^{\dagger }\delta d+{\delta d}^{\dagger }\delta 
c\right)
\end{align}
where, ${G=g}_{OM}c_S$ is the enhanced optomechanical coupling strength due to 
the driving optical field and $\Delta'_c=\Delta_c-g_{OM}(b_s+b_s^\dagger)$ is the modified cavity detuning. {Fig. 2 displays 	
the level diagram of the linearized Hamiltonian and all the coupling routes among 
states denoted by $\left|n_c,n_d,n_b\right\rangle$ in the 	displaced frame; 
where $n_c,n_d$ and $n_b$ are the photon, exciton and 	phonon numbers 
respectively. Different kinds of cooling and heating processes 
may occur due to the optomechanical interaction [34]. The cooling processes 
associated with energy swapping and counter-rotating-wave interaction are 
illustrated by solid (red) curves A and B respectively. The heating processes are denoted 
by the dashed (purple) curves, corresponding to swap heating (C) and quantum 
backaction heating (D). The energy swapping due to the extra exciton-cavity coupling is shown by curves E and F. In order to 
achieve efficient mechanical motion cooling, one needs to enhance the cooling 
effect while suppressing the heating. %The cavity field has a very high decay rate as compared to the high-Q exciton 
%mode. Hence $|2\rangle$ represents the short-lived state with decay rate $k_c$ 
%while $|3\rangle$ represents the long-lived metastable state with decay rate 
%$k_d$. Destructive quantum interference between the two paths, namely  
%$|1\rangle \to |2\rangle$ and $|1\rangle\to |2\rangle\to |3\rangle\to 
%|2\rangle$ suppresses the heating process and results in cooling.
}The Langevin equations for the fluctuation terms in the linearized Hamiltonian are given by:
\begin{subequations}
\begin{align}\label{eq5}
%\nonumber
\dot{\delta c}=\left(i{\mathrm{\Delta 
	}}_c'-\frac{k_c}{2}\right)\delta 
	c-iG\left(\delta b^{\dagger }+\delta b\right)-igd
	-\sqrt{k_c}c_{in}(t) \\
%	\nonumber
	\dot{\delta d}=\left(i{\mathrm{\Delta }}_d-
	\frac{k_d}{2}\right)\delta d-ig\delta 
	c-\sqrt{k_d}d_{in}(t)\\   
	%\nonumber                  
	\dot{\delta b}=\left(-i{\omega }_m-\frac{{\gamma 
		}_m}{2}\right)\delta b-i{G(\delta c}^{\dagger 
	}+\delta c)-\sqrt{{\gamma}_m}b_{in}(t)  	
\end{align}
\end{subequations}
Eqs. (5a)-(5c) can be solved in the frequency domain to obtain the expression for  
	$\widetilde{\delta b}\left(\omega \right)$ as follows:

	\begin{align}\label{eq6}
	%\nonumber 
	\widetilde{\delta b}\left(\omega \right)=\frac{\sqrt{\gamma_m}\widetilde{b_{in}}\left(\omega \right)-i\sqrt{k_c}A\left(\omega\right)-\sqrt{k_d}B\left(\omega \right)}{i\omega 
		-i\left[\omega_m+\mathrm\Sigma\left(\omega \right)\right]-\frac{{\gamma 
			}_m}{2}} 
	 \\   \nonumber 
	 	\end{align}
	 where, 
	              \begin{align}  
	A\left(\omega \right)=G[\chi \left(\omega 
	\right)\widetilde{c_{in}}\left(\omega \right)+\chi^*\left(-\omega 
	\right)\widetilde{c_{in}}^\dagger\left(\omega \right)]  \nonumber 
	\\	\nonumber
	B\left(\omega\right) =gG[\chi \left(\omega \right)\chi_d\left(\omega 
	\right)\widetilde{d_{in}}\left(\omega \right)-\chi^*\left(-\omega\right)
	\\ \nonumber
	\chi^*_d\left(-\omega \right)\widetilde{d_{in}}^\dagger \left(\omega \right)]   	\
	\nonumber
	\end{align}

%%%%%%%%%%%%%% FIGURE 4 %%%%%%%%%%%
\begin{figure*}[t]
	\centering
	\includegraphics [width =0.45\linewidth,height=6.0 cm]{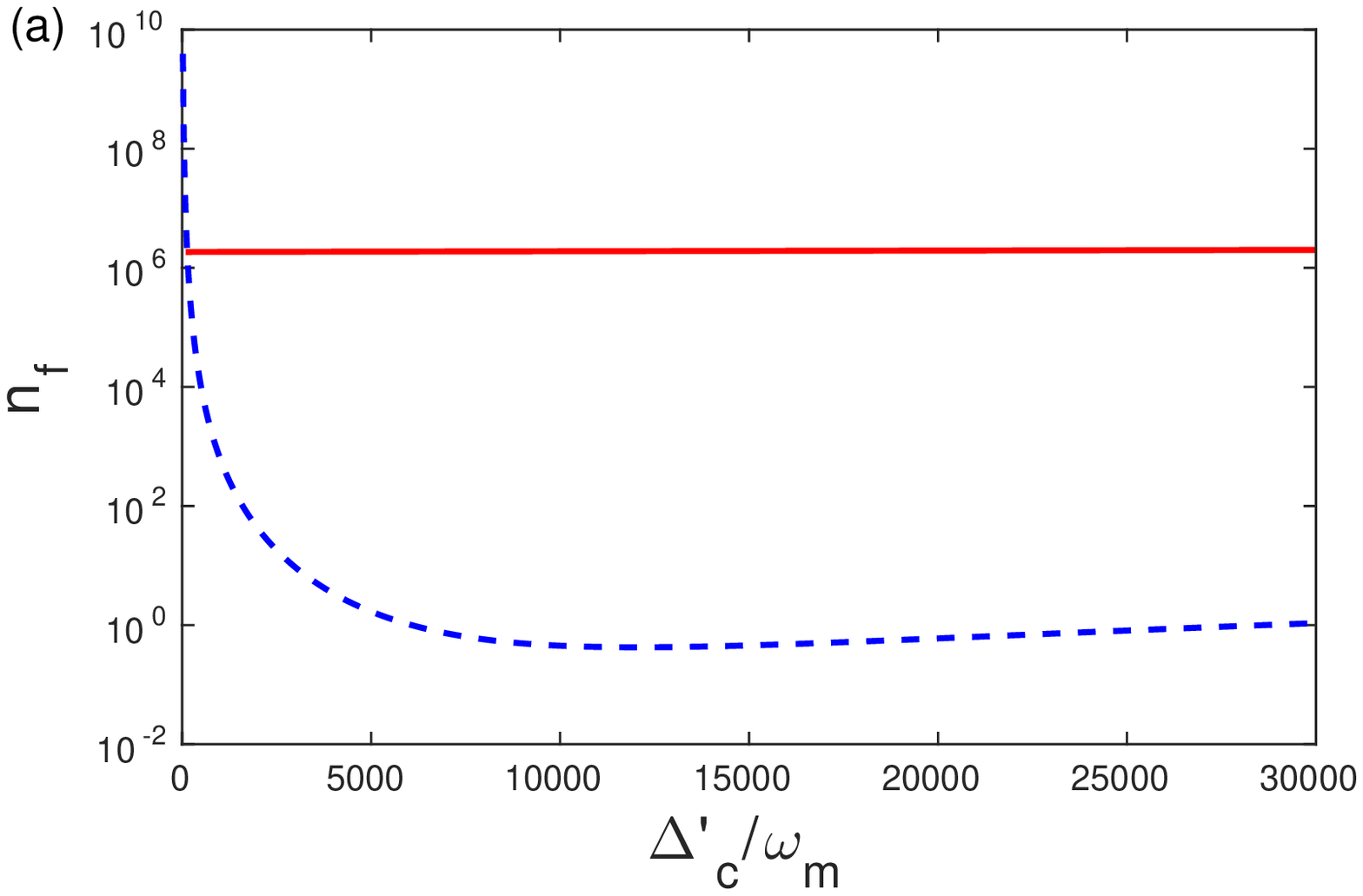}
	\includegraphics [width =0.45\linewidth,height=6.0 cm]{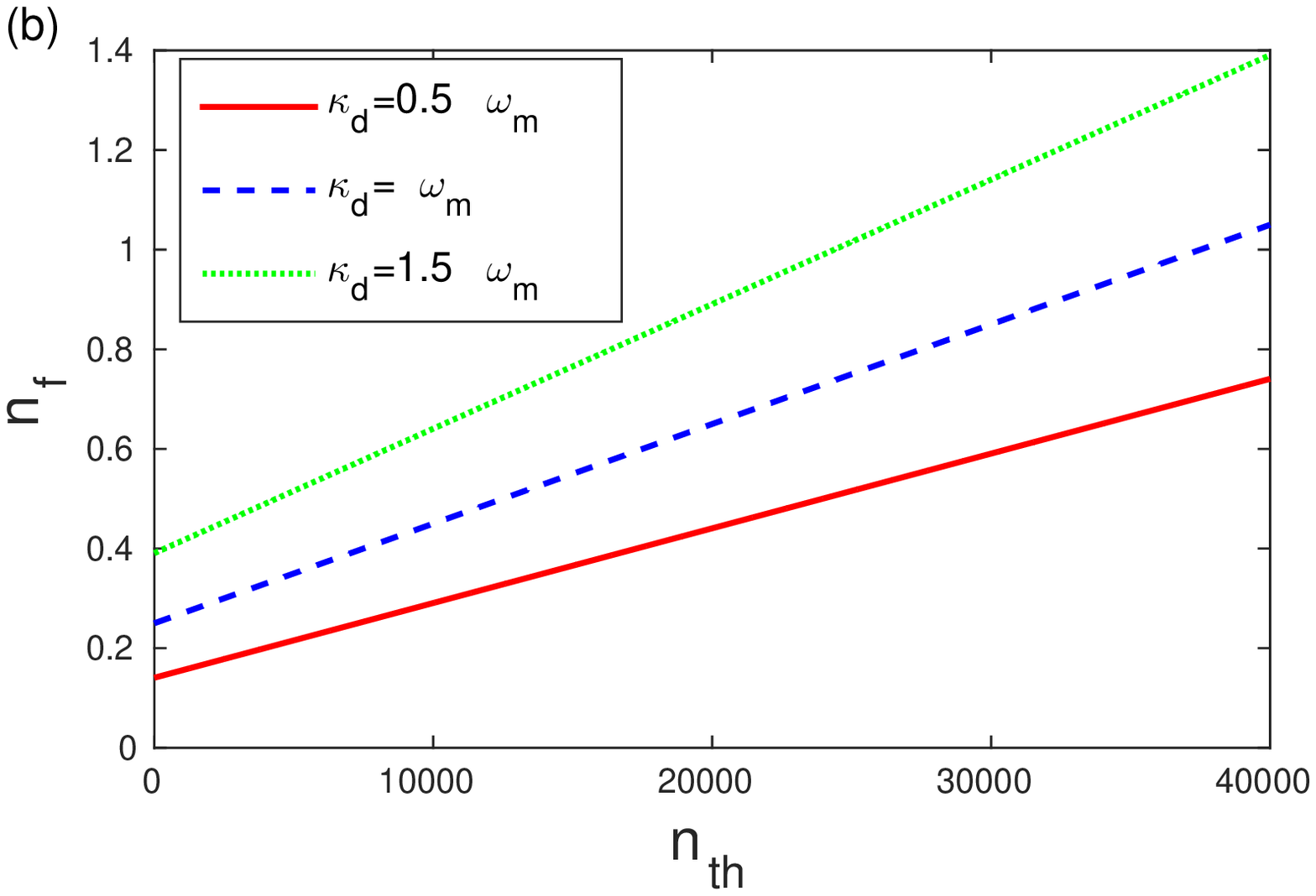}
	\caption {(Color online) Plot of steady-state cooling limit as function of (a) $\Delta_c'/\omega_m$, with $\kappa_d=\omega_m$ and $n_{th}=10^4$. The red solid line denotes the final phonon number in the mechanical oscillator for a generic optomechanical cavity, whereas the blue dotted line shows the phonon number in presence of the QW in the cavity; (b) $n_{th}$ with $\Delta_c'=10^4\omega_m$. Other unspecified parameters are: $\gamma_m=10^{-5}\omega_m$, $k_c=10^4\omega_m$, $g=100\omega_m$, $G=50\omega_m$, $\Delta_d=0.5\omega_m$.}
	\label {fig4}
\end{figure*}
Here, $\mathrm{\Sigma }\left(\omega \right)=-iG^2[\chi \left(\omega 
\right)-{\chi }^*\left(-\omega \right)]$ is the optomechanical self-energy, 
where    $\chi \left(\omega \right)={\left[{\left\{{\chi }_c\left(\omega 
\right)\right\}}^{-1}+g^2{\chi }_d\left(\omega \right)\right]}^{-1}$ is the 
total response 
function of the optomechanical cavity with the QW. ${\chi }_c\left(\omega 
\right)={\left[-i\left(\omega +{\mathrm{\Delta 
}'}_c\right)+\frac{k_c}{2}\right]}^{-1}$, ${\chi }_d\left(\omega 
\right)={\left[-i\left(\omega +{\mathrm{\Delta 
}}_d\right)+\frac{k_d}{2}\right]}^{-1}$ and ${\chi }_m\left(\omega 
\right)={\left[-i\left(\omega -{\omega }_m\right)+\frac{{\gamma 
}_m}{2}\right]}^{-1}$ are the response functions of the optical mode, the 
exciton mode and the mechanical mode respectively. The radiation pressure force, 
in an optomechanical system, arising due to the interaction term, $H_{int}$ , is 
given by $F=-\frac{\delta H_{int}}{\delta x}$. Using this, the radiation 
pressure force for our system is estimated as  $F=-G\ [{\delta c}^{\dagger 
}+\delta c]/x_{ZPF}$ , where, $x_{ZPF}$ is the zero-point fluctuation of the 
mechanical oscillator.  The quantum noise spectrum is calculated 
using: $S_{FF}\left(\omega \right)=\int{dt e^{i\omega t}\left\langle F\ 
	(t)F(0)\right\rangle }$ [6]. The spectral density, in our system is calculated 
to be:
\begin{align}\label{eq7}
	S_{FF}\left(\omega \right)=\frac{G^2{\left|\chi \left(\omega 
			\right)\right|}^2}{x^2_{ZPF}}\left[k_c+k_dg^2{\left|{\chi 
		}_d\left(\omega 
		\right)\right|}^2\right]
\end{align}
The cooling rate of the mechanical resonator is given by 
$A_-=S_{FF}\left({\omega }_m\right)x^2_{ZPF}$ while the heating rate is given by 
$A_+=S_{FF}\left(-{\omega }_m\right)x^2_{ZPF}$. Due to the dynamical back-action induced by the radiation 
pressure force, the spring constant (and thereby the effective oscillation 
frequency) and the damping rate of the mechanical oscillator get modified. The 
extra damping of the mechanical oscillator due to the optomechanical interaction 
is given by, ${\mathrm{\gamma}}_{\mathrm{OM}}=A_--A_+=-2Im\mathrm{\Sigma}({\omega }_m)$ 
and the mechanical frequency shift is given by, $\delta {\omega 
}_m=Re\mathrm{\Sigma }({\omega }_m)$. In Fig. 3, we have plotted the 
noise spectrum for different values of modified cavity-field detuning in the 
unresolved-sideband regime. {It is worthwhile to mention that in our calculations 
we have used realistic parameters inspired from a recent experimental work [36].} 
In a generic optomechanical cavity, in 
the unresolved-sideband regime, the noise spectrum reduces to 
$S_{FF}\left(\omega \right)=\frac{G^2{\left|\chi \left(\omega 
\right)\right|}^2}{x^2_{ZPF}}k_c$. This gives rise to a Lorentzian curve 
illustrating equal heating and cooling rates of the mechanical 
resonator.Nevertheless, the noise spectrum for the system considered here as 
depicted in Fig. 3, shows asymmetric Fano lineshapes that arise as a result of 
interference between resonant and nonresonant processes [23]. 
{This indicates that the presence of the QW modulates the cavity profile to show asymmetry in cooling and heating processes. For $\Delta_c'=k_c$, there is an increase in cooling rate while the heating rate is reduced, and for $\Delta_c'=-k_c$, the opposite happens. Therefore, by 
tuning the cavity-field detuning, the cooling rate of the mechanical resonator 
can potentially be enhanced while reducing the heating rate. This is possible
due to the interaction of the high-Q QW with the mechanical mode through the  
cavity field. Intuitively, the asymmetry in cooling and heating rates can be pictured as a outcome of quantum interference. As can be observed from Fig. 2, due to the presence of the exciton-cavity coupling $g$, there are two different pathways leading to the same cooling or heating process. Therefore, one can harness the interference to overpower the heating effect while enhancing the cooling effect.} 
		
In the highly unresolved regime, $k_c\gg {\omega }_m$ and for $k_c\gg {(k}_d,\ 
{\gamma }_m)$, $\Delta_c'\gg {\Delta_d}$, the three-mode system can be reduced to a two-mode system by 
considering the cavity mode as 	perturbation. Integrating Eqs. (5a)-(5c), we get 
the time dependent form of the operators as follows:
\begin{subequations}
\begin{align}\label{eq8}
\nonumber
 \delta c\left(t\right) =\delta c\left(0 \right)\mathrm{exp} \left(i{\mathrm{\Delta}}'_ct-\frac{k_c}{2}t\right) + 
 \mathrm{exp}\left(i\Delta'_ct-\frac{k_c}{2}t\right) &
 \\ \nonumber
\int_0^t[-iG\delta b^{\dagger}\left(\tau \right)-
iG\delta b\left(\tau \right)
-ig\delta 
d(\tau)-\sqrt{k_c}c_{in}(\tau )] \times &
\\ %\nonumber
\mathrm{exp} \left(-i{\mathrm{\Delta }}'_c\tau +\frac{k_c}{2}\tau \right) d\tau  &
\\ \nonumber
\delta d\left(t\right) =\delta d\left(0\right){\mathrm{exp}\left(i{\mathrm{\Delta }}_dt-\frac{k_d}{2}t\right)}+{\mathrm{exp}} 
\left(i{\mathrm{\Delta }}_dt-\frac{k_d}{2}t\right) &
\\ %\nonumber
\int^t_0[-ig\delta c(\tau )-
\sqrt{k_d}d_{in}(\tau )]\mathrm{exp}
\left(-i{\mathrm{\Delta}}_d\tau + \frac{k_d}{2}\tau \right) d\tau &
\\ \nonumber
\delta b\left(t\right) =\delta b\left(0\right)\exp \left(-i{\omega }_mt-\frac{\gamma_m}{2}t\right)+
\exp \left(-i\omega_mt-\frac{\gamma_m}{2}t\right) &
\\ \nonumber
\int^t_0[-iG\delta c^\dagger\left(\tau\right)-iG\delta c(\tau )
-\sqrt{\gamma_m}b_{in}(\tau)] &
\\ %\nonumber 
\exp \left(i\omega_m\tau +\frac{{\gamma}_m}{2}\tau \right)d\tau 
\end{align}
\end{subequations}
Now considering the effect of the cavity mode 
as perturbation, 
			the expressions for the time dependent exciton mode and 
mechanical mode operators are approximated to be as follows:
\begin{subequations}
\begin{align}\label{eq9}
%\nonumber
\delta d\left(t\right)\cong \delta 
			d\left(0\right){\mathrm{exp} \left(i{\mathrm{\Delta 
}}_dt-
				\frac{k_d}{2}t\right)}+D_{in}(t)       
\\			%\nonumber
\delta b\left(t\right)\cong \delta b\left(0\right){\mathrm{exp} \left(-i{\omega }_mt-\frac{{\gamma 
					}_m}{2}t\right)\ }+B_{in}(t)   
\end{align}
\end{subequations}
% % % % % % % % % % % % % % % % FIGURE 5 % % % % % % %
\begin{figure*}[t]
	\centering
	\includegraphics [width =0.45\linewidth,height=6.0 cm]{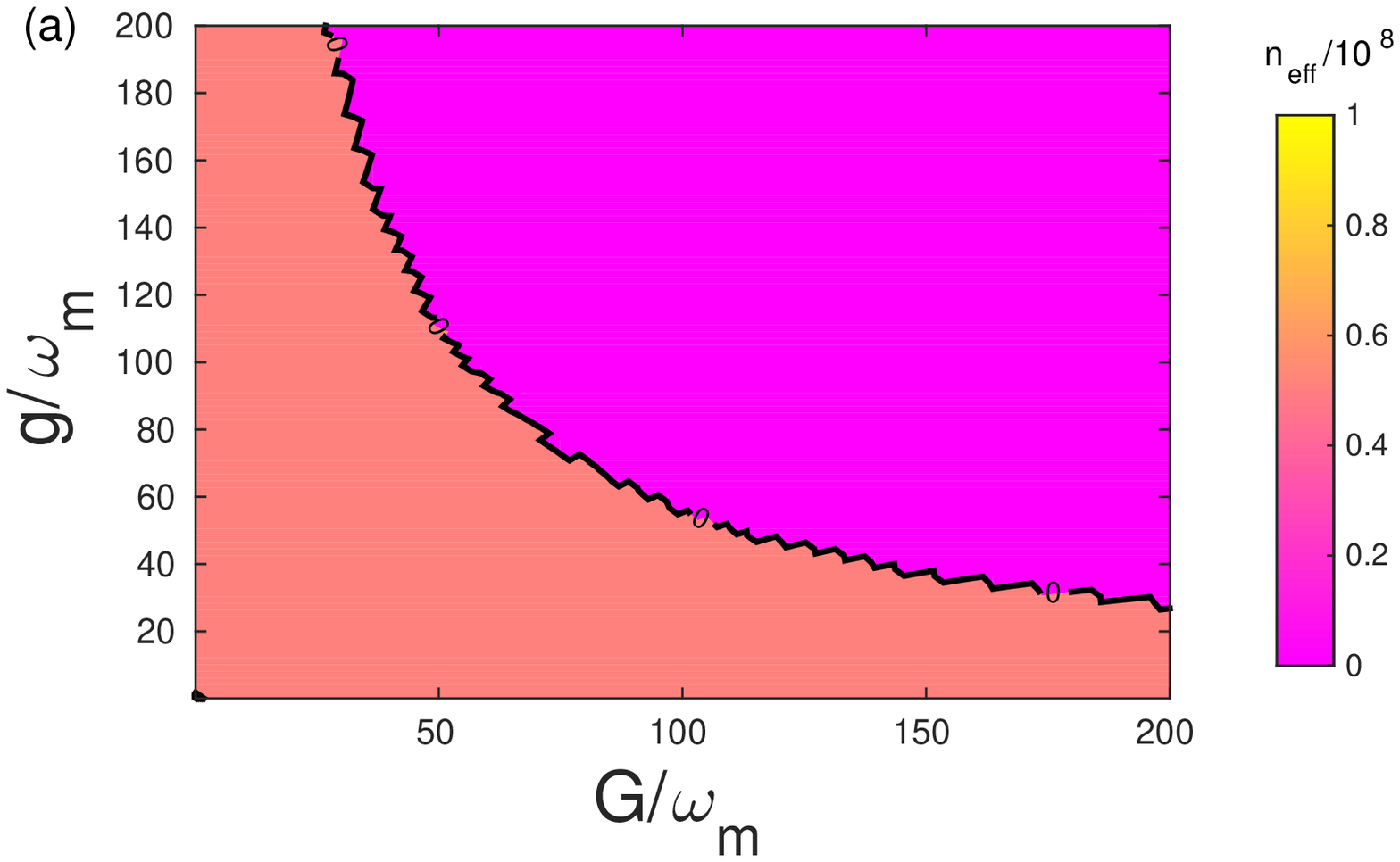}
	\includegraphics [width =0.45\linewidth,height=6.0 cm]{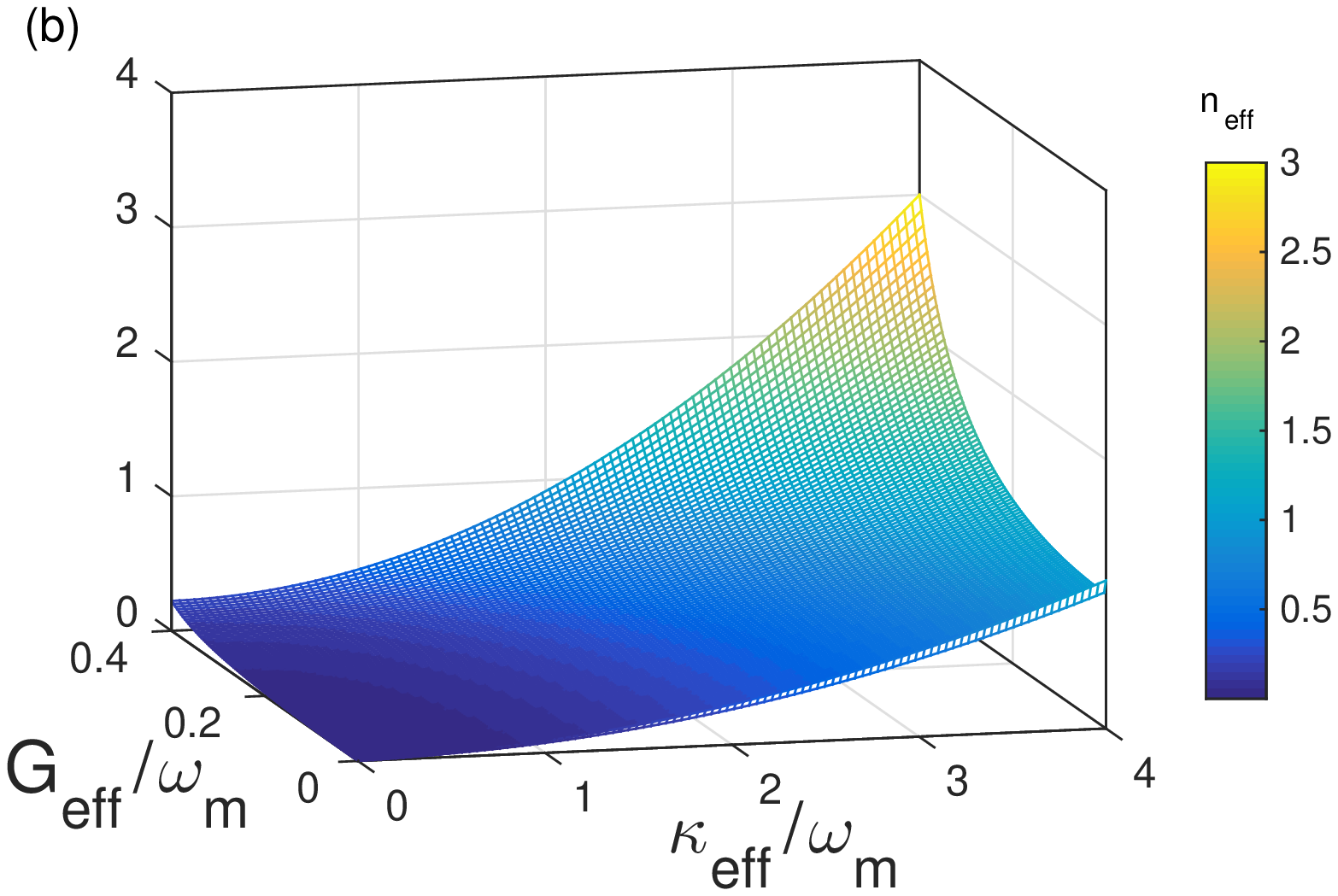}
	\caption {(Color online) (a) Contour plot of mechanical steady-state cooling limit as function of normalized optomechanical coupling $G$ and the exciton-cavity coupling $g$. Other parameters used are: $\gamma_m=10^{-5}$$\omega_m$, $k_c=10^4\omega_m$, $k_d=\omega_m$, $n_{th}=10^4$, $\Delta_c'=10^4\omega_m$ and  $\Delta_d=0.5\omega_m$. (b) Plot of phonon occupancy limit as a function of normalized effective parameters $k_{eff}$ and $G_{eff}$  for  $\gamma_m n_{th}=10^{-9} \omega_m$.}
	\label {fig5}
\end{figure*}
where the effect of the cavity mode is included in the noise terms $D_{in}(t)$ and $B_{in}(t)$ . Substituting Eqs. (9a) and (9b) back 
			into Eq. (8a) and under the assumptions 
$\left|{{\mathrm{\Delta }}'}_c \right|\gg {\mathrm{\Delta }}_d,k_c\gg (k_d,\gamma_m)$ , we obtain: 

\begin{align} \label{eq10} 
\nonumber
\delta c\left(t\right) \cong -\frac{iG\left[\delta b\left(t\right)+\delta b^\dagger \left(t\right)\right]}{-i\Delta'_c+\frac{k_c}{2}}- \frac{ig\delta 					
d\left(t\right)}{-i\Delta'_c+\frac{k_c}{2}}+
\\ \delta c\left(0\right)\exp \left(i\Delta'_ct-\frac{k_c}{2}t\right)+C_{in}(t) 
\end{align} 
				And substituting $\delta c\left(t\right)\ 
$into Eqs. (5b) and (5c), we get the equation for the 
exciton mode as:
\begin{align} \label{eq11} 
\nonumber 
\dot{\delta d}=\left(i\Delta_d-\frac{k_d}{2}\right)\delta d	
\\ \nonumber	
	+ig\left[\frac{iG\left[\delta b\left(t\right)+{\delta b}^{\dagger}\left(t\right)\right]}{-i\Delta'_c+\frac{k_c}{2}} 
 +\frac{ig\delta d\left(t\right)}{-i\Delta'_c+\frac{k_c}{2}}\right]
 \\ 
 -\sqrt{k_d}d_{in}\left(t\right)-
  ig[\delta c\left(0\right) \exp \left(i\Delta_c't-\frac{k_c}{2}t\right)+
 C_{in}(t)] 
\end{align} 
Comparing this with the generic single-cavity optomechanical system, we 
can derive the parameters for the effective exciton-mechanical mode 
interaction as
 ${\mathrm{\Delta }}_{eff}={\mathrm{\Delta }}_d-{\eta 
						}^2{\mathrm{\Delta }}_c'$, 
$k_{eff}=k_d+{\eta }^2k_c$, $G_{eff}=\eta G$, 
where $\eta =g/\sqrt{{\mathrm{\Delta}}^{'2}_c+{\left(\frac{k_c}{2}\right)}^2}$ that can be approximated to $\eta =g/\Delta_c'$ for   $\Delta_c'\gg k_c$. Analogous to the single 
							cavity optomechanical 
system, the steady-state cooling limits are 
							approximated as 
$n_{eff}=n_{classical}+n_{quantum}$ [34, 37]. 					
Here,						
$n_{classical}=\frac{4G^2_{eff}+k^2_{eff}}{4G^2_{eff}k_{eff}}
\gamma_mn_{th} \approx \frac{k_{eff}}{4{\left|G_{eff}\right|}^2}$$\gamma_mn_{th}$ is the classical steady-state cooling limit and $n_{quantum}=\frac{k^2_{eff}+8G^2_{eff}}{16({\omega }^2_m-4G^2_{eff})}\ 
\approx \frac{k^2_{eff}}{16{\omega }^2_m}$   is the quantum limit of 
								cooling for 
effective resolved sideband $(k_{eff}\ll {\omega }_m)$ and 
								the effective 
weak coupling regime ($G_{eff}<k_{eff}$). In Figs. 4(a)-4(b) the variations of
steady-state cooling limit as function of normalized cavity detuning and the thermal phonon number are shown. As 
seen from Fig. 4(a), ground state cooling is not possible for a 
generic optomechanical cavity in the unresolved-sideband regime. 
Nevertheless, in case of the QW coupled system, ground state cooling can be 
achieved for a range of high cavity detuning near ${10}^4\omega_m$. {For large detuning, dark modes with respect to the low-Q cavity mode ($c$) are formed via linear combination of the mechanical mode ($b$) and high-Q exciton mode ($d$). This dark-mode is responsible for an effective cooling [32]. It is worth to be noted that in case of a single optomechanical cavity without the QW, cooling occurs in the red detuned regime only. But, with the QW inside the cavity, the damping is significantly enhanced in the blue-detuned regime as illustrated in Fig. 3. The effective detuning term $\Delta_{eff}$ contains both $\Delta_c'$ and $\Delta_d$. Hence, it is possible to tune these blue-detuned terms to get a red-detuned $\Delta_{eff}$ at high value of $\Delta_c'$, that results in cavity cooling. Fig. 4(b) depicts the variation of the steady-state  cooling limit as a function of the bath phonon number for different values of $k_d$. The plots show that, for ground state cooling of the mechanical resonator, high values of bath phonon number is tolerable. For example, for $k_d={\omega }_m$, the maximum tolerable bath phonon number is approximately equal to ${37\mathrm{\times }10}^3$. It is also to be noted that more bath phonon 
number is tolerable for ground state cooling with lower decay rate of the 
QW excitons.

It is important to have an idea about the optimum range of the optomechanical coupling $G$ and the exciton-cavity coupling $g$ to be used for efficient mechanical mode cooling. These couplings are fixed by the material at the fabrication stage. Though the optomechanical coupling $G$ also depends on the input laser power, it is difficult to tune the couplings at a later stage. Fig. 5(a) demonstrates the optimal range of $g$ and $G$. The optimum value of $g$ can be found out analytically by maximizing the cooling rate, that is found to be $g=\sqrt{(\frac{\Delta_c'+\omega_m}{\Delta_d+\omega_m})(\frac{k_d}{2})^2 +(\Delta_c'+\omega_m)(\Delta_d+\omega_m)}$. For the stability of the system, the effective coupling must follow $G_{eff}<{\omega }_m/2$, as observed from the quantum limit of cooling. Therefore, the condition for the optomechanical coupling $G$ and the exciton-cavity coupling $g$ 
simplifies to $gG<{\omega }_m\sqrt{{\mathrm{\Delta}}^{'2}_c+{\left(\frac{k_c}{2}\right)}^2}/2$ . For ${\mathrm{\Delta }}'_c\cong {\kappa }_c$, the requirement for the approximate value of the couplings give $gG<{\omega}_m{\kappa }_c/2$. In Fig. 5(b), the variation 
of the steady-state cooling limit as function of effective parameters 
$k_{eff}$ and $G_{eff}$, is shown. It illustrates that ground state cooling of mechanical motion may be achieved under the stability condition for $G_{eff}$ and for the range of $k_{eff}<4\omega_m$, as indicated by the quantum limit of cooling.}

In order to study the time evolution of the mean phonon number in the mechanical resonator, we use the master equation approach. The quantum master equation of the system reads:			
%%%%%%%%%%%%%% FIGURE 6 %%%%%%%%%%%
\begin {figure}[t]
\begin {center}
\includegraphics [width =9 cm]{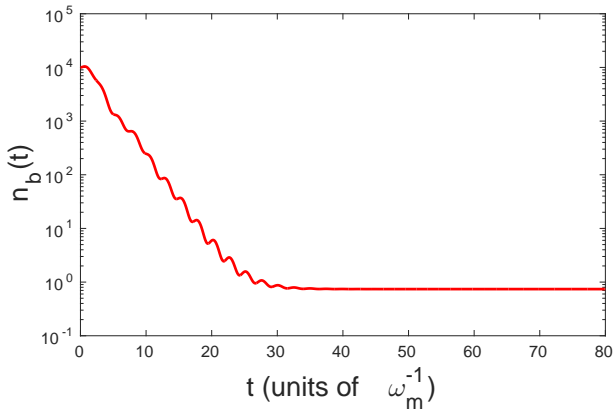}
\caption{Color online) Time evolution of the mean phonon number in the mechanical resonator in presence of the QW, starting from $n_{th}=10^4$. Parameters considered are $k_c=10^4\omega_m$, $k_d=\omega_m$, $\Delta_c'=10^4 \omega_m$, $\Delta_d=0.5\omega_m$, $\gamma_m=10^{-5}\omega_m$, $g=100\omega_m$ and $G=50\omega_m$.}
\label{fig6}
\end{center}
\end{figure}								
\begin{align} \label{eq12} 
\nonumber							
\dot{\rho}=i\left[\rho ,H_L\right]+\frac{k_c}{2}\left(2\delta c\rho 		
{\delta c}^{\dagger }-{\delta c}^{\dagger }\delta c\rho -\rho 
\delta c^{\dagger }\delta c\right)
\\ \nonumber
+\frac{k_d}{2}\left(2\delta d\rho {\delta d}^{\dagger }-{\delta d}^{\dagger }\delta d\rho -\rho {\delta d}^{\dagger }\delta d\right)
\\ \nonumber
+\frac{{\gamma }_m}{2}\left(n_{th}+1\right)\left(2\delta 
b\rho {\delta b}^{\dagger }-\delta b^{\dagger }\delta b\rho -\rho {\delta 
b}^{\dagger }\delta b\right)
\\ 
+\frac{{\gamma}_m}{2}n_{th}(2\delta b^{\dagger }\rho \delta b-\delta b\delta b^{\dagger}\rho -\rho \delta b{\delta b}^{\dagger })											
\end{align} 								
We use the covariance approach to find out the time evolution of the mean 
phonon number $n_b\left(t\right)=\langle\delta b^{\dagger }\delta b\rangle (t)$ 
[34]. For this, we solve a linear system of differential equations
${\partial }_t\left\langle {\hat{o}}_i{\hat{o}}_j\right\rangle =Tr\left(\dot{\rho }{\hat{o}}_i{\hat{o}}_j\right)=\sum_{m,n}{{\mu}_{m,n}\left\langle {\hat{o}}_m{\hat{o}}_n\right\rangle }$, where, ${\hat{o}}_i$, ${\hat{o}}_j$, ${\hat{o}}_m$, ${\hat{o}}_n$ are one of the 								
operators: ${\delta b}^{\dagger }$, $\delta c^{\dagger }$, ${\delta 	d}^{\dagger }$, $\delta b$, $\delta c$ and $\delta d$; and ${\mu }_{m,n}$ 		
are the corresponding coefficients.  Solving these, we can find out the 	
mean values of all the time-dependent second-order moments: $\langle\delta b^{\dagger }\delta b\rangle $, $\langle {\delta b}^{\dagger}\delta c\rangle $, $\langle {\delta b}^{\dagger }\delta d\rangle $, $\langle {\delta c}^{\dagger }\delta c\rangle $, $\langle \delta c^{\dagger }\delta d\rangle $, $\langle \delta d^{\dagger }\delta d\rangle $,  $\langle {\delta b}^2\rangle $,$\ \langle \delta b\delta c\rangle $, $\langle \delta b\delta d\rangle $, $\langle {\delta c}^2\rangle $,  $\langle \delta c\delta d\rangle $ and  $\langle \delta d^2\rangle $. In Fig. 6 we show the time evolution of the mean phonon number. The environmental phonon number is assumed to be ${10}^4$. The cavity is considered to be in highly unresolved sideband regime. Initially the phonon number in the mechanical oscillator is equal to the environmental phonon number. All other second order moments are initially 
zero. The plot shows that with increasing time, the average phonon occupancy in the mechanical resonator is cooled down to below 1. This indicates ground state cooling of the resonator mode in the highly unresolved sideband regime.
\section{\label{sec:level1}CONCLUSION}										
 In conclusion, we have studied the sideband cooling of a 
mechanical resonator in an optomechanical cavity containing a quantum 
well. It is worthwhile to note that such semiconductor structures, with 
well-developed semiconductor fabrication techniques, are easily 
integrable with cavities and waveguides making them a unique tool for 
exploiting optomechanics. The exciton and mechanical modes are not 
coupled directly, but their interaction with the cavity optical 
field gives rise to an indirect coupling between them. This specific 
configuration of the system can lead to cooling of the mechanical 
oscillator in the unresolved sideband regime. Due to the presence of the high-Q element in the cavity, the noise spectrum is modified and leads to asymmetric cooling and heating rates. Even when the cavity 
is in the highly unresolved-sideband regime, the effective interaction between the 
exciton and mechanical modes can bring the system back to effective 
resolved-sideband regime. Hence the requirement of the resolved-sideband
condition for cooling is relaxed significantly. The cooling rate of the 
mechanical resonator can be enhanced by tuning the cavity-field detuning.
The time evolution of the mean phonon number in the mechanical resonator 
is studied using the quantum master equation. It is found that, with 
increasing time, the average phonon occupancy in the mechanical resonator
tends towards zero exhibiting dynamic controllability of cavity 
dissipation. This might open up the possibility of manipulation of 
semiconductor integrated mechanical systems in the quantum mechanical 
regime.
\begin{flushleft}
\textbf{Acknowledgements}
\end{flushleft}
 B. Sarma would like to thank MHRD, Government of India for a research fellowship.
\\\\
\noindent \textbf{References}
\\\\
\noindent [1] C. H. Metzger and K. Karrai, Nature (London) \textbf{432},
1002 (2004).\\
\noindent [2] D. Kleckner and D. Bouwmeester, Nature (London) 
\textbf{444}, 75 (2006).	\\				
\noindent [3] T. Corbitt, Y. Chen, E. Innerhofer, H. M\"{u}ller-Ebhardt, 
D. Ottaway, H. Rehbein, D. Sigg, S. Whitcomb, C. Wipf, and N. Mavalvala, 
Phys. Rev. Lett. \textbf{98}, 150802 (2007).\\
\noindent [4] A. Schliesser, O. Arcizet, R. Rivi\`{e}re, G. Anetsberger 
and T. J. Kippenberg, Nat. Phys. \textbf{5}, 509 (2009).\\
\noindent [5] S. Gr\"{o}blacher, J. B. Hertzberg, M. R. Vanner, G. D. 
Cole, S. Gigan, K. C. Schwab \& M. Aspelmeyer, Nat. Phys. \textbf{5}, 485 
(2009).									\\
\noindent [6] M. Aspelmeyer, T. J. Kippenberg, and F. Marquardt, Rev. 
Mod. Phys. \textbf{86}, 1391, (2014).	\\
\noindent [7] V. B. Braginsky and A. B. Manukin, Sov. Phys. JETP
\textbf{25}, 653 (1967).			\\	
\noindent [8] V. B. Braginsky and V. S. Nazarenko, Sov. Phys. JETP 
\textbf{30}, 770 (1970).				\\	
\noindent  [9]V\textit{. }B. Braginsky, A. B. Manukin, and M. Yu. 
Tikhonov, Sov. Phys. JETP   \textbf{31}, 821 (1970).\\
\noindent  [10]  C. Fabre, M. Pinard, S. Bourzeix, A. Heidmann, E. 
Giacobino, and S. Reynaud, Phys. Rev. A\textbf{ 49, }1337 (1994).\\
\noindent [11] Samuel Aldana, Christoph Bruder, and Andreas Nunnenkamp, 
Phys. Rev. A \textbf{88}, 043826 (2013).	\\
\noindent [12] V. B. Braginsky, \textit{Measurement of weak forces in 
Physics Experiments} (University of Chicago Press, Chicago, 1977).\\
\noindent [13] G. Anetsberger, O. Arcizet, Q. P. Unterreithmeier, R.
Rivi\`{e}re, A. Schliesser, E.M. Weig, J. P. Kotthaus, and T. J. 
Kippenberg, Nat. Phys. \textbf{5}, 909 (2009); J. D. Teufel, R. Donner,
M. A. Castellanos-Beltran, J.W. Harlow, and K.W. Lehnert, Nat. 
Nanotechnology, \textbf{4}, 820 (2009); A. A. Clerk, M. H. Devoret, S. M. 
Girvin, F. Marquardt, and R. J. Schoelkopf, Rev. Mod. Phys. \textbf{82}, 
1155 (2010); T. P. Purdy, R.W. Peterson, and C. A. Regal, Science 
\textbf{339}, 801 (2013).		\\		
\noindent [14] D. Vitali, S. Gigan, A. Ferreira, H. R. B\"{o}hm, P.
Tombesi, A. Guerreiro, V. Vedral, A. Zeilinger, and M. Aspelmeyer, Phys.
Rev. Lett. \textbf{98}, 030405 (2007); C. Genes, A. Mari, P. Tombesi, and 
D. Vitali, Phys. Rev. A \textbf{78}, 032316 (2008); Y.-D. Wang and A. A. 
Clerk, Phys. Rev. Lett. \textbf{110}, 253601 (2013).\\
\noindent [15] S. Mancini, V. I. Man'ko and P. Tombesi, Phys. Rev. A, 
\textbf{55}, 3042 (1997); S. Bose, K. Jacobs, and P. L. Knight, Phys. 
Rev. A \textbf{56, }4175 (1997); M. Paternostro, Phys. Rev. Lett. 
\textbf{106}, 183601 (2011).	\\		
\noindent [16] L. Tian and H. L. Wang, Phys. Rev. A \textbf{82}, 053806
(2010); L. Tian, Phys. Rev. Lett. \textbf{108}, 153604 (2012); C. Dong,
V. Fiore, M. C. Kuzyk, H. Wang, Science \textbf{338}, 1609 (2012); T. A. 
Palomaki, J. W. Harlow, J. D. Teufel, R. W. Simmonds and K. W. Lehnert,
Nature (London) \textbf{495}, 210 (2013).	\\
\noindent [17] G. S. Agarwal and S. Huang, Phys. Rev. A \textbf{81}, 
041803 (2010); S. Weis, R. Riviere, S. Deleglise, E. Gavartin, O. 
Arcizet, A. Schliesser, and T. J. Kippenberg, Science \textbf{330}, 1520 
(2010); A.H. Safavi-Naeini, T. P.Mayer Alegre, J.Chan, M. Eichenfield, 
M.Winger, Q. Lin, J. T. Hill, D. E. Chang, andO. Painter, Nature (London) 
\textbf{472}, 69 (2011); H. Wang, X. Gu, Y. Liu, A. Miranowicz, and F. 
Nori, Phys. Rev. A \textbf{90}, 023817 (2014).\\
\noindent [18] A. D. O'Connell, M. Hofheinz, M. Ansmann, R. C. Bialczak, 
M. Lenander, E. Lucero, M. Neeley, D. Sank, H. Wang, M. Weides, J. 
Wenner, J.M. Martinis, and A. N.Cleland, Nature (London) \textbf{464}, 
697 (2010).					\\		
\noindent [19] Florian Marquardt, Joe P. Chen, A. A. Clerk, and S. M. 
Girvin, Phys. Rev. Lett. \textbf{99}, 093902 (2007); \\
\noindent [20] J. D. Teufel, T. Donner, D. Li, J. W. Harlow, M. S. 
Allman, K. Cicak, A. J. Sirois, J. D. Whittaker, K.W. Lehnert, and R.W. 
Simmonds, Nature (London) \textbf{475}, 359 (2011); J. Chan, T. P. 
Alegre, A. H. Safavi-Naeini, J. T. Hill, A. Krause, S. Groeblacher, 
M.Aspelmeyer, and O. Painter, Nature (London) \textbf{478}, 89 (2011); A.
H. Safavi-Naeini, J. Chan, J. T. Hill, T. P. Mayer Alegre, A. Krause, and 
O. Painter, Phys. Rev. Lett. \textbf{108}, 033602 (2012).\\
\noindent [21] D. J. Wineland and W. M. Itano, Phys. Rev. A \textbf{20}, 
1521 (1979).			\\				
\noindent [22] I. Wilson-Rae, N. Nooshi, W. Zwerger, and T. J.
Kippenberg, Phys. Rev. Lett. \textbf{99}, 093901 (2007); J. D. Teufel, J. 
W. Harlow, C. A. Regal, and K. W. Lehnert, Phys. Rev. Lett. \textbf{101}, 
197203 (2008). 		\\					
\noindent [23] F. Elste, S. M. Girvin, and A. A. Clerk, Phys. Rev. Lett. 
\textbf{102, }207209 (2009).	\\		
\noindent [24] M. Li, W. H. P. Pernice, and H. X. Tang, Phys. Rev. Lett.
\textbf{103}, 223901 (2009); A. Xuereb, R. Schnabel, and K. Hammerer, 
Phys. Rev. Lett. \textbf{107}, 213604 (2011); T. Weiss and A. Nunnenkamp,
Phys. Rev. A \textbf{88}, 023850 (2013); M.-Y. Yan, H.-K. Li, Y.-C. 
Liu,W.-L. Jin, and Y.-F. Xiao, Phys. Rev. A \textbf{88}, 023802 (2013).\\
\noindent [25] W.-J. Gu and G.-X. Li, Phys. Rev. A \textbf{87}, 025804 
(2013); Y. Guo, K. Li, W. Nie, and Y. Li, Phys. Rev. A \textbf{90}, 
053841  (2014); Y.-C. Liu, Y.-F. Xiao, X. Luan, Q. Gong, and C.W.Wong, 
Phys. Rev. A \textbf{91}, 033818 (2015).\\
\noindent [26] C. Genes, H. Ritsch, and D. Vitali, Phys. Rev. A 
\textbf{80}, 061803(R) (2009); K. Hammerer, K. Stannigel, C. Genes, P. 
Zoller, P. Treutlein, S. Camerer, D. Hunger, and T. W. H\"{a}nsch, Phys.
Rev. A \textbf{82}, 021803(R) (2010); C. Genes, H. Ritsch, M. Drewsen, 
and A. Dantan, Phys. Rev. A \textbf{84}, 051801(R) (2011); S. Camerer, M.
Korppi, A. J\"{o}ckel, D. Hunger, T. W. H\"{a}nsch, and P. Treutlein, 
Phys. Rev. Lett. \textbf{107}, 223001 (2011); B. Vogell, K. Stannigel, P.
Zoller, K. Hammerer, M. T. Rakher, M. Korppi, A. J\"{o}ckel, and P. 
Treutlein, Phys. Rev. A \textbf{87}, 023816 (2013); F. Bariani, S. Singh,
L. F. Buchmann, M. Vengalattore, and P. Meystre, Phys. Rev. A 
\textbf{90}, 033838 (2014); A. Dantan, B. Nair, G. Pupillo, and C. Genes,
Phys. Rev. A \textbf{90}, 033820 (2014); J. S. Bennett, L. S. Madsen, M. 
Baker, H. Rubinsztein-Dunlop, and W. P. Bowen, New J. Phys. \textbf{16},
083036 (2014); G. Ranjit, C. Montoya, and A. A. Geraci, Phys. Rev. A 
\textbf{91}, 013416 (2015); T. P. Purdy, D. W. C. Brooks, T. Botter, N.
Brahms, Z.-Y. Ma, and D. M. Stamper-Kurn, Phys. Rev. Lett. \textbf{105},
133602 (2010); X. Chen, Y.-C. Liu,  P. Peng, Y. Zhi,  and Y.-F. Xiao, 
Phys.  Rev. A \textbf{92}, 033841 (2015).\\
\noindent [27] T. Ojanen and K. B{\o}rkje, Phys. Rev. A \textbf{90}, 
013824 (2014); Y.-C. Liu, Y.-F. Xiao, X. S. Luan, and C. W. Wong, Sci. 
China Phys. Mech. Astron. \textbf{58}, 050305 (2015).\\
\noindent [28] K. Hennessy,  A. Badolato, M. Winger, D. Gerace, M. 
Atat\"{u}re, S. Gulde, S. F\"{a}lt, E. L. Hu, and A. Imamo\^{u}glu, 
Nature \textbf{445,} 896 (2007); M. Nomura, N. Kumagai, S. Iwamoto, Y. 
Ota, and Y. Arakawa, Nature Phys., \textbf{6,} 279 (2010); S. Hughes and 
H. J. Carmichael, New J. Phys. \textbf{15}, 053039 (2013).\\
\noindent [29] L. Ding, C. Baker, P. Senellart, A. Lemaitre, S. Ducci, G.
Leo, and I. Favero, Phys. Rev. Lett. \textbf{105}, 263903 (2010); L.
Ding, C. Baker, P. Senellart, A. Lemaitre, S. Ducci, G. Leo, and I. 
Favero, Appl. Phys. Lett. \textbf{98}, 113108 (2011); K. Usami, A. 
Naesby, T. Bagci, B. M. Nielsen, J. Liu, S. Stobbe, P. Lodahl, and E. S. 
Polzik, Nat. Phys. \textbf{8}, 168 (2012); A. Fainstein, N. D. 
Lanzillotti-Kimura, B. Jusserand, and B. Perrin, Phys. Rev. Lett. 
\textbf{110}, 037403 (2013); S. Anguiano, G. Rozas, A. E. Bruchhausen, A.
Fainstein, B. Jusserand, P. Senellart, and A. Lemaitre, Phys. Rev. B 
\textbf{90}, 045314 (2014).\\
\noindent [30] E. A. Sete and H. Eleuch, Phys. Rev. A \textbf{85}, 043824 
(2012); O.    Kyriienko, T. C. H. Liew, and I. A. Shelykh, Phys. Rev.
Lett. \textbf{112}, 076402 (2014). \\
\noindent [31] E. A. Sete, H. Eleuch, and C. H. Raymond Ooi, Phys. Rev. A 
92, 033843 (2015).\\
\noindent [32] Y-D Wang and A. A. Clerk, Phys. Rev. Lett. \textbf{108}, 
153603 (2012); C. Dong, V. Fiore, M. C. Kuzyk and H. Wang, Science 
\textbf{338}, 1609 (2012); S. A. McGee, D. Meiser, C. A. Regal, K.W. 
Lehnert and M. J. Holland, Phys. Rev. A 87, 053818 (2013).\\
\noindent [33] K. Bergmann, H. Theuer, and B. W. Shore, Rev. Mod. Phys.
\textbf{70}, 1003 (1998). \\
\noindent [34] Y. Liu, Y. Xiao, X. Luan and C. W. Wong, Phys. Rev. Lett. 
~\textbf{110}, 153606 (2013). \\
\noindent [35] C.W. Gardiner and P. Zoller, \textit{Quantum Noise 
}(Springer-Verlag, Berlin, 1991); B. Peng, S. K. \"{O}zdemir, F. Lei, F. 
Monifi, M. Gianfreda, G. L. Long, S. Fan, F. Nori, C. M. Bender and L. 
Yang, Nat. Phys. \textbf{10}, 394 (2014). \\
\noindent [36] A. Fainstein, N. D. Lanzillotti-Kimura, B. Jusserand, and 
B. Perrin, Phys. Rev. Lett. ~\textbf{110}, 037403 (2013).\\
\noindent [37] J. M. Dobrindt, I. Wilson-Rae, and T. J. Kippenberg, Phys. 
Rev. Lett. \textbf{101}, 263602 (2008).

\end{document}